\documentclass{article}
\usepackage[final,nonatbib]{nips_2017}
\usepackage[numbers]{natbib}
\usepackage[utf8]{inputenc} % allow utf-8 input
\usepackage[T1]{fontenc}    % use 8-bit T1 fonts
\usepackage{url}            % simple URL typesetting
\usepackage{booktabs}       % professional-quality tables
\usepackage{amsfonts}       % blackboard math symbols
\usepackage{nicefrac}       % compact symbols for 1/2, etc.
\usepackage{microtype}      % microtypography
\usepackage{graphicx,float}
\usepackage{amssymb,amsmath,amsthm}
\usepackage{cleveref}
\usepackage{multirow,sparklines}
\usepackage{soul}
\usepackage{hyperref}       % hyperlinks
\newcommand{\ignore}[1]{}

\title{A multiobjective deep learning approach \\ for predictive classification in Neuroblastoma}
\author{
\begin{tabular}{c}
Valerio Maggio, Marco Chierici, Giuseppe Jurman\textsuperscript{*}, Cesare Furlanello\\
Fondazione Bruno Kessler, Trento (IT)\\
\textsuperscript{*} Corresponding author: \texttt{jurman@fbk.eu}
\end{tabular}
}

\begin{document}
\maketitle
\begin{abstract}
Neuroblastoma is a strongly heterogeneous cancer with very diverse clinical courses that may vary from spontaneous regression to fatal progression;
an accurate patient's risk estimation at diagnosis is essential to design appropriate tumor treatment strategies.
Neuroblastoma is a paradigm disease where different diagnostic and prognostic endpoints should be predicted from common molecular and clinical information, with increasing complexity, as shown in the  FDA MAQC-II study.
Here we introduce the novel multiobjective deep learning architecture CDRP (Concatenated Diagnostic Relapse Prognostic) composed by 8 layers to obtain a combined diagnostic and prognostic prediction from high-throughput transcriptomics data.
Two distinct loss functions are optimized for the Event-Free Survival (EFS) and Overall Survival (OS) prognosis, respectively.
We use the High-Risk (HR) diagnostic information as an additional input generated by an autoencoder embedding.
The latter is used as network regulariser, based on a clinical algorithm commonly adopted for stratifying patients from cancer stage, age at insurgence of disease, and MYCN, the specific molecular marker.
The architecture was applied to Illumina HiSeq2000 RNA sequencing of 498 neuroblastoma patients (176 at high risk) from the Sequencing Quality Control (SEQC) study, obtaining state-of-art on the diagnostic endpoint and improving prediction of prognosis over the HR cohort.
\end{abstract}
\paragraph{Introduction}
The challenge of dealing with multiple endpoints of clinical interest is a hallmark of predictive models from high-throughput molecular data, as demonstrated in the MAQC-II (Microarray Analysis and Quality Control) Study \cite{maqc10maqcII}. Neuroblastoma is a paradigmatic example of disease where the medical community has adopted a clinical algorithm that defines the subtype of cancer patients with lowest expectation of response to therapy and survival, but the precision medicine approach is still failing to identify molecular profiles clearly associated to patient subtypes. Especially for High risk (HR) patients, adequate therapies are still lacking.

Arising predominantly in the first two years of life, neuroblastoma is the most frequent extracranial solid tumor in infancy, accounting for about 500 new cases in Europe per year (130 in Germany), corresponding to roughly 8\% of pediatric cancers and 15\% of pediatric oncology deaths~\cite{maris07neuroblastoma}.
Neuroblastoma develops from the immature cells of the ganglionic sympathetic nervous system lineage stemming from the neural crest cells, and tumors can arise at any site where sympathetic neuroblasts are present during normal development~\cite{mohlin13hif2a}, \textit{e.g.}, in chest.
The broad variety of clinical behavior represent neuroblastoma's major hallmark, ranging from spontaneous regression (stage 4S) to gradual maturation (stages 1-2) to aggressive and often fatal ganglioneuroma~\cite{ambros09international,rozmus12multiple} (stages 3-4), despite intensive multimodal treatment. Official staging is defined by the International Neuroblastoma Staging System (INSS)~\cite{brodeur1993revisions}.
The current strategies used to appropriately design tumor treatment therapies use different combinations of clinical and genetic markers to discriminate patients with low or high risk of death from disease.
The markers used in this diagnosis include age~\cite{london05evidence}, tumor stage~\cite{evans71proposed,brodeur93revisions} and MYCN proto-oncogene genomic amplification~\cite{brodeur84amplification,seeger85association}. %, as shown in the decision tree of Fig.~\ref{fig:HR}.
However, this standard protocol is still imperfect, often resulting in over- or undertreatment of patients with neuroblastoma~\cite{oberthuer15revised}.
Cancer genetic instability is most often studied at the genomic and gene expression levels, focusing on the effects of genomic alterations on transcription and splicing.
In fact, several studies demonstrated that using messenger RNA (mRNA) expression information for molecular classification improves the diagnostic accuracy over traditional clinical markers for individual tumor behavior, enhancing the risk stratification reliability and therefore the therapy selection~\cite{ohira05expression,asgharzadeh06prognostic,oberthuer06customized,vermeulen09predicting,depreter10accurate,oberthuer10prognostic,maqc10maqcII,formicola16gene}.
Only a limited number of the published classifiers based on gene expression have been so far incorporated into clinical operative systems for a controlled validation trial: as examples,~\cite{saulnier12pilot,stricker14validation} and the U.S. National Institutes of Health clinical trials \cite{children16studying,children16gene}.
The reasons are diverse and include logistic and bureaucratic hindrances for the implementation of classifiers into clinical practice, difficulties in the setup of controlled validation trials for relatively small patient numbers, and the challenge to appropriately design the therapy according to genomic classification results.
Moreover, prognostic gene expression signatures for neuroblastoma stemming from different methodologies applied to different datasets often identify diverse gene sets~\cite{shohet12redefining,valentijn12functional}.
Thus, the impact of genomic classification-induced treatment on the outcome of neuroblastoma patients is still an open issue.
As a contribute, we present here a novel multi-objective deep learning~\cite{lecun15deep} solution named CDRP (Concatenated Diagnostic Relapse Prognostic) that accurately classifies patients in a internationally collected neuroblastoma cohort, by combining both prognostic and diagnostic information from gene expression data.
An artificial neural network (multilayer perceptron) has been used for neuroblastoma outcome prediction~\cite{cangelosi16artificial} from expression data but in a shallow learning framework. Deep learning based approaches have also appeared in the neuroblastoma literature, but using images rather than omics inputs~\cite{salazar17neuroblastoma}.

\begin{figure}
\includegraphics[width=0.95\textwidth]{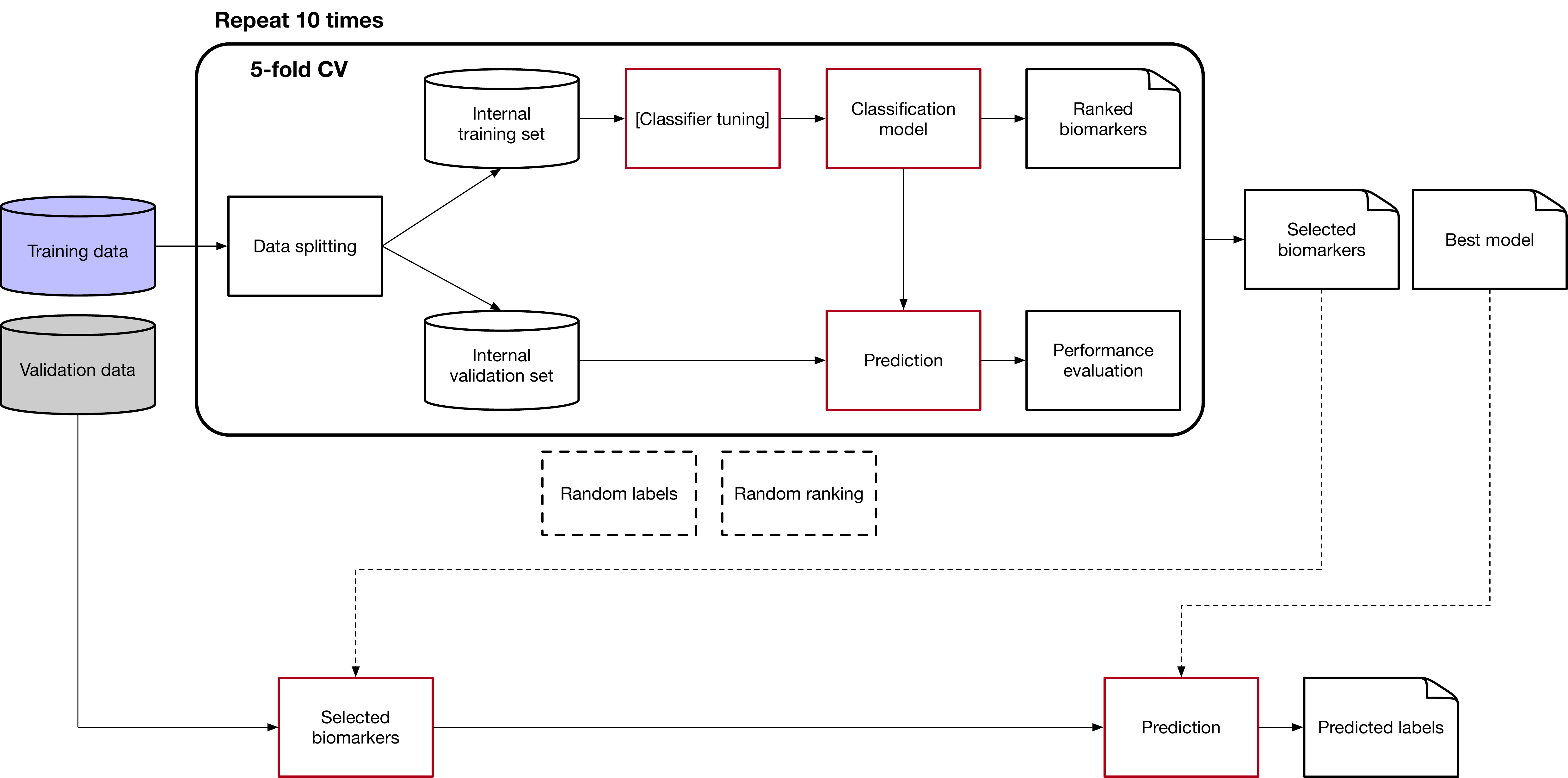}
\caption{The Data Analysis Protocol (DAP) used in the experiments, originally defined in the US-FDA MAQC-II initiative.}
\label{fig:dap}
\end{figure}

Our architecture is built in multiple steps. We train on half of the patients a multitask net CDRP-$\mathcal{N}$ for classification over two distinct prognostic tasks at 5-years, namely Event-Free Survival (EFS: events are relapse, disease progression or death), and the Overall Survival (OS: partitioning patients as either dead or alive).
Furthermore, the shared layer of the multitask net has additional inputs from another network modeling the high-risk (HR: high risk, non high-risk or unknown status) endpoint. In detail, we link values from the embedding of an autoencoder CDRP-$\mathcal{A}$ developed over the same training data for the HR diagnostic task.
In order to control for selection bias, both the net CDRP-$\mathcal{N}$ and the autoencoder CDRP-$\mathcal{A}$ are trained and evaluated using a Data Analysis Protocol (DAP), based on a 10$\times$5-fold cross validation developed within the MAQC-II and SEQC studies led by the US-FDA~\cite{maqc10maqcII,seqc14comprehensive}.
We apply the CDRP-$\mathcal{N}$--CDRP-$\mathcal{A}$ architecture on the RNA sequencing (RNA-Seq) dataset from the SEQC study~\cite{seqc14comprehensive,zhang15comparison}. The same dataset split employed in the Neuroblastoma SEQC satellite study is adopted for comparability~\cite{zhang15comparison}.
We obtain consistent or improved performance with CDRP-$\mathcal{N}$ with respect to SVM or Random Forest models.
Operatively, the framework is implemented in a Keras~\cite{chollet15keras} environment over a Tensorflow~\cite{abadi15tensorflow} backend, run on a nVidia Pascal-GPU Blade equipped with two GTX 1080, 8 GB dedicated RAM, 2560 CUDA cores, up to 9TFlops throughput and 8 CPU Intel Core i7–6700 with 32 GB RAM.
\paragraph{Data description}
The dataset used in this study collects RNA-Seq gene expression profiles of 498 neuroblastoma patients, published as part of the SEQC initiative \cite{seqc14comprehensive,zhang15comparison}.
We considered the following endpoints for classification tasks: the occurrence of an event (progression, relapse or death) (Event-Free survival, ``EFS''); the occurrence of death from disease (Overall Survival, ``OS''); the occurrence of an event (``$\textrm{EFS}_\textrm{HR}$'') and death from disease (``$\textrm{OS}_\textrm{HR}$'') in high-risk (HR) patients only.
HR status was defined according to the NB2004 risk stratification criteria.
The samples were split into training (NBt) and validation (NBv) sets following a published partitioning \cite{zhang15comparison}.
Stratification statistics for NBt and NBv are reported in Tab.~\ref{tab:tr_val}.
RNA-Seq data were preprocessed as $log_2$ normalized expressions for 60,778 genes (``MAV-G'')~\cite{zhang15comparison}.
Expression tables were filtered before downstream analyses by removing features without EntrezID and with interquartile range (IQR) $>0.5$ using the \textit{nsFilter} function in the \textit{genefilter} R package, leaving 12,464 (20.5\%) genes for downstream analysis.
To avoid information leakage, feature filtering was performed on NBt data set and applied on both NBt and NBv sets.
\begin{table}[!t]
\begin{center}
\caption{Sample stratification (left) and summary statistics (right) for the NBt and NBv subset for the covariates High-Risk (HR), Overall Survival (OS) and Event-Free Survival (EFS). HR 0:non high risk/NA (green), 1:high risk (red), EFS 0:no event (green), 1:event (red), OS 0:alive (green), 1:dead (red).}
\label{tab:tr_val}
\begin{tabular}{|r|r|r||r|r|}
\hline
HR & EFS & OS & NBt & NBv \\
\hline
\multirow{3}{*}{0} & 0 & 0 & 129 & 130 \\
\cline{2-5}
& \multirow{2}{*}{1} & 0 & 26  & 24\\
\cline{3-5}
&&1& 8 & 5\\
\hline
\multirow{3}{*}{1} & 0 & 0 & 31 & 25\\
\cline{2-5}
&\multirow{2}{*}{1} & 0 & 12 & 16 \\
\cline{3-5}
&&1& 43 & 49\\
\hline
\end{tabular}
\hspace{1cm}
\begin{tabular}{ccccc}
&&0& 1\\
\multirow{2}{*}{HR} & NBt & 163 & 86 &
\begin{sparkline}{9.82}
\definecolor{sparkrectanglecolor}{named}{olive}
\sparkrectangle 0 1
\end{sparkline}
\begin{sparkline}{5.18}
\definecolor{sparkrectanglecolor}{named}{red}
\sparkrectangle 0 1
\end{sparkline}\\
& NBv & 159 & 90 &
\begin{sparkline}{9.58}
\definecolor{sparkrectanglecolor}{named}{green}
\sparkrectangle 0 1
\end{sparkline}
\begin{sparkline}{5.42}
\definecolor{sparkrectanglecolor}{named}{orange}
\sparkrectangle 0 1
\end{sparkline}\\
\\
\multirow{2}{*}{EFS} & NBt & 160 & 89 &
\begin{sparkline}{9.64}
\definecolor{sparkrectanglecolor}{named}{olive}
\sparkrectangle 0 1
\end{sparkline}
\begin{sparkline}{5.36}
\definecolor{sparkrectanglecolor}{named}{red}
\sparkrectangle 0 1
\end{sparkline}\\
& NBv & 155 & 94 &
\begin{sparkline}{9.34}
\definecolor{sparkrectanglecolor}{named}{green}
\sparkrectangle 0 1
\end{sparkline}
\begin{sparkline}{5.66}
\definecolor{sparkrectanglecolor}{named}{orange}
\sparkrectangle 0 1
\end{sparkline}\\
\\
\multirow{2}{*}{OS} & NBt & 198 & 51 &
\begin{sparkline}{11.93}
\definecolor{sparkrectanglecolor}{named}{olive}
\sparkrectangle 0 1
\end{sparkline}
\begin{sparkline}{3.07}
\definecolor{sparkrectanglecolor}{named}{red}
\sparkrectangle 0 1
\end{sparkline}\\
& NBv & 195 & 54 &
\begin{sparkline}{11.75}
\definecolor{sparkrectanglecolor}{named}{green}
\sparkrectangle 0 1
\end{sparkline}
\begin{sparkline}{3.25}
\definecolor{sparkrectanglecolor}{named}{orange}
\sparkrectangle 0 1
\end{sparkline}\\
\end{tabular}
\end{center}
\end{table}
\paragraph{The deep learning architecture}
\begin{figure}[!b]
\centering
\includegraphics[width=0.7\textwidth]{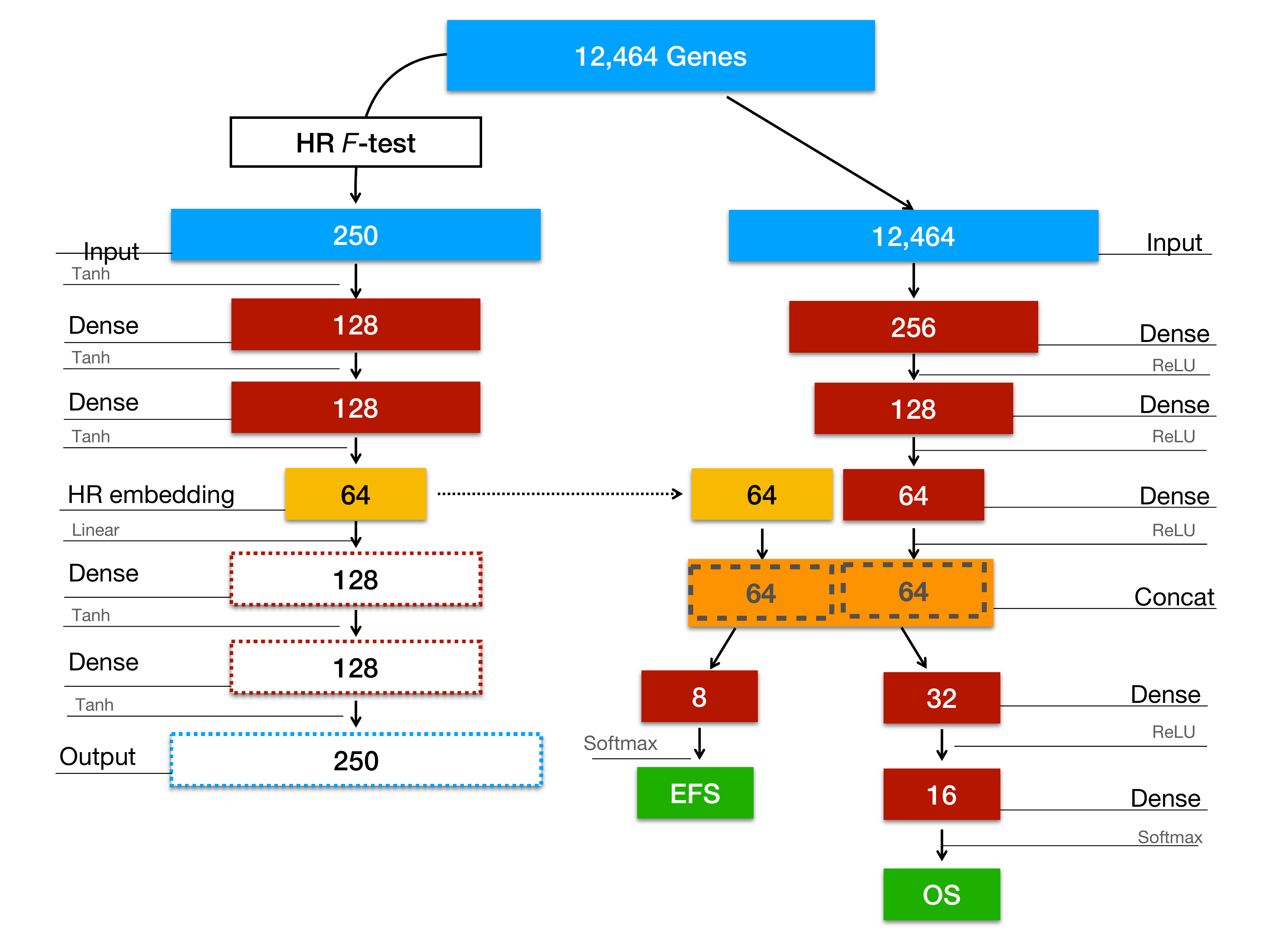}
\caption{The layer/node structure of the proposed deep learning architecture CDRP.}
\label{fig:the_net}
\end{figure}
The architecture of two deep learning solutions CDRP-$\mathcal{N}$ and CDRP-$\mathcal{A}$ are shown in Fig.~\ref{fig:the_net}.
For both, the neural network is developed within a DAP described in detail in the next section.
The autoencoder CDRP-$\mathcal{A}$ is used as a regressor on the HR/non-HR task, aimed at minimizing the mean square error $\textit{mse}$.
The input layer is selected by the DAP K-best algorithm to be of dimension 250 (2\% of the total number of features), corresponding to the best value $\textit{mse}=0.042$ with confidence interval $(0.041;0.043)$.
This is followed by two dense layers with 128 nodes with $\tanh$ activation, ending in another dense layer with 64 nodes with linear activation.
The output is later used as the HR embedding input for the shared merge layer in CDRP-$\mathcal{N}$.
A specular decoding structure (dotted boxes and arrows in Fig.~\ref{fig:the_net}) exists but is not used by CDRP-$\mathcal{N}$.
The classification net CDRP-$\mathcal{N}$ starts with an initial layer taking as input the whole set of 12,646 features.
This is followed by two dense layers, with 256 and 128 nodes, respectively.
The output of this last layer is merged with concatenation with the HR embedding layer computed by CDRP-$\mathcal{A}$, obtaining a shared layer from which two different branches depart.
Up to this layer, all activations are LeakyReLU function~\cite{nair10rectified,maas14rectifier} with coefficient 0.3, with no dropout~\cite{srivastava14dropout} nor batch normalization~\cite{ioffe15batch}.
The first branch consists of a single output layer with 8 nodes and softmax activation on the EFS task, while the second branch has two layers, first a 32-node dense one and then a 16-ndoe dense softmax activated output layer for the OS task.
The loss function for the OS task has weight 2, while for EFS the weight is 1.
Across all CDRP-$\mathcal{N}$ the batch size is 64, and the optimizer is Adadelta~\cite{zeiler12adadelta} with $\delta=0$ and $\eta=1$.
Finally, for CDRP-$\mathcal{N}$ the number of epochs is bounded to 500, with an early stopping rule on the validation loss, with patience 4 and $\min_\Delta$ $10^{-6}$, while CDRP-$\mathcal{A}$ has 2000 epochs without early stopping.
\paragraph{The analysis pipeline}
The experimental methodology outlined in Fig.~\ref{fig:dap} follows the DAP developed in the context of the MAQC-II challenge~\cite{maqc10maqcII}, the U.S. Food and Drug Administration (FDA) initiative aimed to establish reproducibility in microarray gene expression experiments.
Given a dataset divided in a training and a test set, the former undergoes a $10\times 5-$fold Stratified Cross Validation resulting in ranked list of features and a classification performance measure, here the Matthews Correlation Coefficient~\cite{matthews75comparison,baldi00assessing} $\textrm{MCC} = \frac{\textrm{TP} \cdot \textrm{TN} - \textrm{FP}\cdot\textrm{FN}}{\sqrt{(\textrm{TP}+\textrm{FP})(\textrm{TP}+\textrm{FN})(\textrm{TN}+\textrm{FP})(\textrm{TN}+\textrm{FN})}}$, for TN, TP, FN, FP the entries of the binary confusion matrix.
Data are mean zero and variance one and $\log_2$ transformed before undergoing classification, and in order to avoid information leakage standardization parameters from the training set are used for both training and test subsets.
The $k$-best algorithm is chosen as the feature ranker, and the classification is performed using the deep learning architecture previously described, and the best model is later retrained on the whole training set and selected for validation on the test set.
Furthermore, as a sanity check to avoid unwanted selection bias effects, the pipeline is repeated 20 times with two randomized strategies: a Random Label scheme where the true training labels are stochastically scrambled, and a Random Feature scheme where a random set of features is selected instead of the optimal list.
\paragraph{Results and discussion}
CDRP is a novel multitask deep learning architecture that improves prediction of hard prognostic endpoints by injecting latent variables from autoencoding the standard clinical model.
The performance of the CDRP architecture is summarized in Tab.~\ref{tab:results}.
CDRP improves MCC in validation for the OS endpoint, and it is the first model to improve on the High Risk cohort (EFS-HR, OS-HR).
A theoretical basis justifying the achieved improvement relies on the fact that the information distilled from the diagnostic task adds clinical information, used by the multi-task predictor, which combines the OS and EFS tasks.
Finally, CDRP models with random labels yield $\mathrm{MCC}\approx0$, indicating honest estimates, while consistent results are obtained also with swapped training and validation sets.
Retraining of CDRP-$\mathcal{N}$ on the HR subset is especially computing intensive and will be added in a future version of the paper. Further refinement will include the extraction of enriched pathways for genes derived from the activation in the shared layer.
\begin{table}[!t]
\caption{Comparison of the median MCC from the SEQC study in cross-validation (``NBt'') and external validation (``NBv'') with the MCC obtained by CDRP-$\mathcal{N}$. For LSVM, RF, $\mathcal{N}_s$ and CDRP-$\mathcal{N}$, 95\% studentized bootstrap confidence intervals for NBt are also reported. The chosen architecture CDRP-$\mathcal{N}$ was the best performing in cross-validation on NBt. MCC values for CDRP-$\mathcal{N}$ on $\textrm{EFS}_\textrm{HR}$ and $\textrm{OS}_\textrm{HR}$ are inherited from the values on the overall cohort, with no retraining.}
\label{tab:results}
\begin{center}
\resizebox{\textwidth}{!}{%
\begin{tabular}{c|rr|rr|rr|rr|rr}
Task & \multicolumn{2}{c|}{SEQC} & \multicolumn{2}{c|}{LSVM} & \multicolumn{2}{c|}{RF} & \multicolumn{2}{c}{$\mathcal{N}_s$} & \multicolumn{2}{c}{CDRP-$\mathcal{N}$} \\
 & NBt & NBv & NBt & NBv & NBt & NBv & NBt & NBv & NBt & NBv \\
\hline
EFS & 0.45 & 0.50 & 0.46 (0.43;0.49) & 0.48 & 0.45 (0.41;0.48) & 0.52                        & 0.40 (0.36;0.45) & 0.41 & 0.42 (0.38;0.45) & 0.45 \\
OS & 0.48 & 0.47 & 0.46 (0.42;0.50) & 0.47 & 0.43 (0.39;0.47) & 0.37                         & 0.48 (0.46;0.53) & 0.48 & 0.50 (0.45;0.54) & 0.57 \\
$\textrm{EFS}_\textrm{HR}$ & 0.34 & 0.16 & 0.13 (0.08;0.18) & 0.21 & 0.17 (0.10;0.23) & 0.13 & 0.15 (0.09;0.22) & 0.19 & 0.18 (0.11;0.25) & 0.38 \\
$\textrm{OS}_\textrm{HR}$ & 0.36 & 0.07 & 0.22 (0.16;0.28) & 0.12 & 0.33 (0.26;0.39) & 0.10  & 0.23 (0.21;0.35) & 0.14 & 0.25 (0.19;0.31) & 0.19\\
\end{tabular}%
}
\end{center}
\end{table}
%
%%% For camera-ready only
%\subsubsection*{Acknowledgments}
%
\bibliographystyle{unsrtnat}
\small
\bibliography{maggio17multiobjective}
\end{document}